\documentclass{article}

\usepackage{microtype}
\usepackage{graphicx}
\usepackage{subfigure}
\usepackage{booktabs} 

\usepackage{hyperref}

\usepackage[accepted]{icml2025}

\usepackage{amsmath}
\usepackage{amssymb}
\usepackage{mathtools}
\usepackage{amsthm}

\usepackage[capitalize,noabbrev]{cleveref}

\theoremstyle{plain}

\theoremstyle{definition}

\theoremstyle{remark}

\usepackage[textsize=tiny]{todonotes}

\icmltitlerunning{Improved Receiver Chain Performance via Error Location Inference}

\begin{document}

\twocolumn[
\icmltitle{Improved Receiver Chain Performance via Error Location Inference}



\icmlsetsymbol{equal}{*}

\begin{icmlauthorlist}
\icmlauthor{Michael Greenwood (Mike.Greenwood@CGI.com)}{}
\icmlauthor{Robert Hunter (Rob.Hunter@CGI.com)}{}
\end{icmlauthorlist}

\icmlcorrespondingauthor{Michael Greenwood}{Mike.Greenwood@cgi.com}
\icmlcorrespondingauthor{Robert Hunter}{Rob.Hunter@cgi.com}

\icmlkeywords{Machine Learning, Satellite Communications}

\vskip 0.3in
]




\begin{abstract}
Modern spacecraft communication systems rely on concatenated error correction schemes, typically combining convolutional and Reed-Solomon (RS) codes. This paper presents a decoder-side method that uses a machine learning model to estimate the likelihood of byte-level corruption in received data frames. These estimates are used to mark erasures prior to RS decoding, enhancing its correction capacity without requiring changes to spacecraft hardware or encoding standards. The approach enables improved data recovery under degraded signal conditions at a gain of 0.3 decibels.
\end{abstract}

\section{Introduction}
Spacecraft communication systems operate under constrained signal-to-noise conditions, particularly in low-Earth orbit and deep-space missions \cite{ccsds2017}. To ensure data integrity, receiver chains typically employ concatenated error correction, combining a convolutional inner code with a Reed-Solomon (RS) outer code \cite{wicker1994}. For satellite communications these codes are standardised by the Consultative Committee for Space Data Systems (CCSDS). A common standard widely in use is the pairing of a Convolution coding scheme using the Viterbi algorithm with an RS Forward Error Correction scheme.

RS codes are valued for their reliability. When operated in their hard-decision configuration, they have an extremely low probability of mis-correction \cite{mceliece1986}. A limited number of errors can be corrected by RS, however this algorithm can correct more errors if it knows in advance the positions of the errors \cite{reed1960}. 

In a satellite communications receiver chain, there is no mechanism for identifying which symbols are likely to be in error, and thus no erasure information is passed to the RS decoder. As a result, its full correction capacity is underutilised. This paper proposes a decoding method that uses a machine learning model to estimate the likelihood of corruption at the byte level. The most probable error locations are then given to the RS decoder, allowing it to deal with an increased number of total errors.

The proposed approach introduces a trade-off. Reducing the number of remaining unmarked errors Reed-Solomon needs to locate and correct slightly increases the odds of achieving zero syndromes with a message different from the original. In doing so, we marginally relax the conservative assumptions that underlie standard RS operation; however, the net result remains a configuration in which the probability of mis-correction remains low while achieving a measurable improvement in decoding performance at lower signal to noise ratios (SNR).

The cost of constructing satellite ground stations increases non-linearly with aperture size due to mechanical and structural requirements \cite{townsend2005}. Enhancing decoding capability through software extends the use of smaller ground antennas and reduces pressure to expand physical infrastructure, offering a cost-effective path to improved link performance.

\section{Related Work}

Deep learning-based approaches have been shown to estimate channel state information and detect symbols in OFDM systems, creating end-to-end learnable receiver chain systems until the output of the decoder \cite{yi2020}, and mitigate interference by using neural networks to minimise distortion and interference \cite{cash2022,sun2021}.

For error correction, soft-decision decoding of RS codes has been an established area of study, with multiple efforts to improve performance by incorporating symbol reliability information \cite{koetter2003,wesemeyer2003,jiang2005,elkhamy2005}. These RS decoding techniques do not involve machine learning directly but are conceptually aligned with ML-based error localization methods in that they exploit confidence-weighted information to guide decoding decisions, and are compatible with the techniques described here.

Buckley et al. proposed a closely related approach in which a neural network is used to predict error regions based on soft outputs from a turbo decoder for a proposed image transmission system \cite{buckley2000}. The predicted regions are used to generate erasure flags for an outer Reed–Solomon (RS) decoder, improving correction performance under channel degradation. 

The approach is explicitly designed for image data, leveraging the strong spatial correlation between adjacent pixels to guide the neural network's inferences. In this context, symbols are inherently structured. In contrast, the method presented here is not constrained by exploitation of a spatial input format and operates on arbitrary byte sequences. 
The work by Buckley et al. also illustrates the applicability of learned erasure prediction in decoder chains incorporating turbo codes, suggesting potential for extension of this method beyond convolutional–RS systems.

\section{Methodology}

Our method augments a conventional convolutional decoder and RS decoder chain with two additional elements: (1) machine learning–based error position inference, and (2) feedback from successfully RS-corrected interleaved blocks into earlier decoding stages, a previously explored optimisation \cite{Sawaguchi2001}. These are used in a coordinated, iterative process that improves the likelihood of successful frame correction.

\subsection{Decoder Error Correction Feedback}

The decoding pipeline begins with a Viterbi convolutional decoder, followed by RS error correction. When one or more interleaved RS blocks are successfully decoded, the known-correct symbol values from those blocks are used to update the input of the convolutional decoder. This is done through a state pinning procedure: bit-level reliability values corresponding to the corrected RS symbols are replaced with fixed values corresponding to maximum confidence.

The convolutional decoding is then re-run using the updated input stream, potentially reducing bit errors across the frame. This feedback mechanism is repeated whenever additional RS blocks are corrected, tightening decoder performance over successive iterations. 

The focus of our work was to take this adaption and combine it with ML–based erasure estimation rather to achieve a gain greater than the sum of the parts. This is possible because improved decoding increases the availability of potential inferences for locating errors and decreases the number of errors that need to be located for a successful RS error correction.

\subsection{Machine Learning–Based Erasure Estimation}

Following each decoding pass, the machine learning model predicts the likelihood of byte-level corruption across the frame and outputs a per-byte corruption probability. Based on these probabilities, the system constructs a ranked list of suspected error locations.

\subsection{Controlled Erasure Marking and RS Decoding}

To avoid premature mis-correction, erasure marking is introduced incrementally. Initially, the top N most probable error locations (by model confidence) are marked as erasures. The RS decoder is then applied. If decoding fails, N is increased up to a predefined upper bound. This threshold ensures that erasure misclassification does not substantially reduce the system’s overall correctness guarantees.

\subsection{Successive Iteration}

The correction procedure is performed iteratively and after each round:
\begin{itemize}
\item Successfully corrected blocks update both decoder input (via state pinning) and model inference.
\item The convolutional decoder is re-applied to the updated frame.
\item The ML model is re-invoked using the latest decoder outputs and known-correct data.
\item Improved erasure probability estimations are used to re-attempt RS correction.
\end{itemize}
The process continues until no further improvement can be made or the frame is fully corrected.

\section{System Model}
\subsection{Model Architecture}
The error location model uses an attention-based architecture designed to process byte sequences \cite{bahdanau2014,vaswani2017}. The model is formulated as a binary classifier, assigning each input byte a probability of being corrupted. The model does not rely on fixed frame sizes or framing delimiters and makes no assumptions about the syntactic structure of the input data, treating the frame as a generic byte sequence.

\subsection{Data Source and Preparation}
The data used for training, validation, and testing was derived from a Mars Express downlink pass recorded on 27 April 2025. Frames from this pass were recovered using the standard decoding pipeline with convolutional and RS correction applied. These decoded frames served as clean ground truth data for model training.

To generate training examples, each clean frame was stripped of its RS parity and subjected to controlled corruption. Each byte in the frame was independently assigned a small probability of being replaced with a randomly chosen 8-bit value. This substitution yielded noisy input frames paired with binary masks indicating which bytes had been modified. These masks were used as ground truth labels for supervised training.

The dataset was split into 80\% training, 10\% validation, and 10\% test subsets without random shuffling. Only the input data and corresponding corruption masks were used during training.

\subsection{Frame Assumptions}
The target communication system is assumed to use a standard concatenated coding structure with a rate-1/2 convolutional encoder followed by an interleaving process and a (255, 223) Reed-Solomon code, consistent with CCSDS standards. While these parameters were used for evaluation, the machine learning model itself is agnostic to specific code rates or interleaving depths, as it does not explicitly model any of the channel coding structure. 

\subsection{Noise Generation}
Noise is introduced synthetically during training as described above. During evaluation, performance is tested under simulated additive white Gaussian noise (AWGN) conditions using a soft-input convolutional decoder.

\section{Results}
The proposed method was evaluated using frames from a Mars Express downlink captured on 27 April 2025, under AWGN channel conditions. The baseline for comparison was a conventional CCSDS-compliant receiver chain comprising convolutional decoding, interleaving reversal, and RS decoding, without error location inference or decoder feedback.

\subsection{Evaluation Metric}
Performance was measured in terms of bit error rate (BER) as a function of SNR. A frame was considered successfully decoded only if all RS blocks were corrected, and the reconstructed frame exactly matched the original. 

\subsection{Observed Gain}

\begin{figure}[ht]
\vskip 0.2in
\begin{center}
\centerline{\includegraphics[width=\columnwidth]{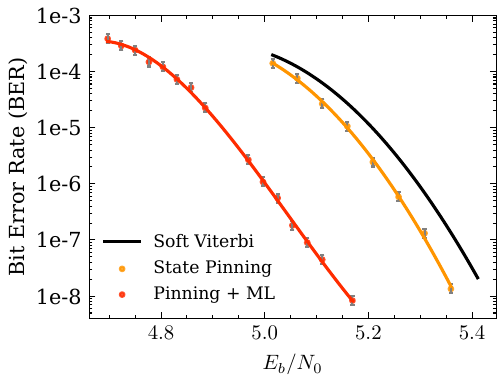}}
\caption{Decibel curves comparing soft Viterbi decoding, the addition of state pinning, and the machine learning implementation with state pinning.}
\label{decibel curves}
\end{center}
\vskip -0.2in
\end{figure}

Across the tested SNR range the integrated system, combining decoder feedback, iterative ML-guided erasure marking, and conservative erasure thresholding achieved a 0.3 dB gain over the baseline at an BER of $10^-6$. 

\subsection{Data Characteristics and Implications}
The frames used for training and evaluation in this study consisted exclusively of payload data generated from the Mars Express missions scientific instruments \cite{esa2004}. In this context, the payload data lacks rigid structure as it reflects continuously varying sensor readings which make it less predictable than other transmitted data types which, by design, exhibit repetitive operational patterns.

The use of payload data therefore represents a conservative test case for this method. More structured or redundant data, such as that found in telemetry channels, would be expected to yield far larger performance improvements under the same methodology. In the satellite communications use case, the recovery value is strongly correlated with high stochasticity payload data; however, in many contexts such as Position, Navigation and Timing data, recovery of a repetitive telemetry signals may be valuable.

This outcome arises from the nature of the machine learning approach. The model infers which bytes are likely to be corrupted based on statistical relationships between bytes within the same frame. These relationships exist because some byte sequences are more probable than others, due to redundancy, protocol constraints, or semantic structure. When the underlying data exhibits stronger or more consistent byte correlations, the model has a richer statistical basis for predicting corruption, which in turn improves erasure marking accuracy and enhances RS decoder performance.

\subsection{Mis-correction Probability}
It is inflexible to fix the trade-off between the probability of a mis-correction and the bit error rate at a single value, and it is expected that different applications of this technology should have different degrees to which trading off the probability of a mis-correction is valuable. For our purposes, the maximum number of erasures marked was set at twenty-two. This leaves up to five errors correctable with unknown location and makes the probability of a mis-correction approximately $10^-12$. The encoding scheme used by MEX also contains two bytes for a cyclic redundancy check, decreasing the probability of a mis-correction further.

\section{Further Work}
Many opportunities exist to extend the current work in three broad categories. The model inputs can be expanded, the output of the model can be modified or differently interpreted, and other changes can be made in the receiver chain which interact constructively with the proposed modelling.

\subsection{Input Improvement}
The training data used in this study was limited to frames recovered from a single MEX downlink pass. While sufficient for demonstrating feasibility, this represents a narrow sampling of the possible channel conditions, frame structures, and encoding practices encountered in operational systems. Expanding the dataset to include telemetry from additional spacecraft, varied mission phases, and alternative modulation or coding schemes would improve the model’s generality and allow for more robust benchmarking.

An additional direction involves expanding the temporal context available to the model. Currently, inference is performed on a single frame in isolation. However, in many spacecraft telemetry systems, successive frames exhibit structured redundancy, shared headers, recurring patterns, or predictable control data sequences.

Inputs can also be qualitatively improved by including probability information in the form of a soft input, or by including data from the demodulator.

\subsection{Output Modifications}
Modelling could treat error correction as a sequence completion task. This would allow the system to propose candidate corrected values directly, potentially improving performance beyond the RS maximum corrections limit. Especially with the introduction of more training data, some bytes can be identified for which there is a high confidence of an error, but also a high confidence of what the byte should have been. A mixture of error location and byte replacement may maximise the opportunity presented by a model which ingests vastly more training data.

\subsection{Complimentary Receiver Chain Optimisations}
There are a wide variety of proposed improvements to the receiver chain, many of which treat the values passed between receiver chain components as probabilities. This paper demonstrates not only the gain from correlations within a frame, but that the combination of machine learning with other approaches improves in a manner that is greater than the sum of the parts. It is the strong recommendation of this paper that future work investigate the value of a combined approach.

Most prominently in this recommendation is the soft-in soft-out (SISO) Reed-Solomon decoders. Algorithms such as Koetter-Vardy decoding accept probabilistic input to further improve gain. 

Future work may also find that with inferences available, the current methods for interleaving are no longer optimal. When frame data becomes easier to recover than parity data by a significant margin, keeping all the parity in a single block at the end of a frame makes burst errors more likely to make a frame unrecoverable. 

\section{Conclusions}
Our work demonstrates a decoder-side machine learning approach for improving Reed-Solomon error correction by identifying likely byte-level corruption and applying targeted erasure marking. Combined with feedback from successfully corrected blocks and iterative state pinning to a convolutional decoder, the method achieved a 0.3 dB gain at a bit error rate of $10^-6$ compared to the equivalent CCSDS-standard decoding chain. Our approach requires no changes to the transmission chain and can be deployed on common hardware.

\bibliography{Improved_Receiver_Chain}

\begin{thebibliography}{17}
\providecommand{\natexlab}[1]{#1}
\providecommand{\url}[1]{\texttt{#1}}
\expandafter\ifx\csname urlstyle\endcsname\relax
  \providecommand{\doi}[1]{doi: #1}\else
  \providecommand{\doi}{doi: \begingroup \urlstyle{rm}\Url}\fi

\bibitem[Bahdanau et~al.(2014)Bahdanau, Cho, and Bengio]{bahdanau2014}
Bahdanau, D., Cho, K., and Bengio, Y.
\newblock Neural machine translation by jointly learning to align and
  translate.
\newblock arXiv preprint arXiv:1409.0473, September 2014.
\newblock URL \url{https://arxiv.org/abs/1409.0473}.

\bibitem[Buckley et~al.(2000)Buckley, Ramos, Hemami, and Wicker]{buckley2000}
Buckley, M.~E., Ramos, M.~G., Hemami, S.~S., and Wicker, S.~B.
\newblock Perceptually-based robust image transmission over wireless channels.
\newblock In \emph{Proceedings 2000 International Conference on Image
  Processing (Cat. No.00CH37101)}, pp.\  128--131 vol.2, Vancouver, BC, Canada,
  2000.
\newblock \doi{10.1109/ICIP.2000.899244}.

\bibitem[Cash(2022)]{cash2022}
Cash, M.~E.
\newblock Neural networks for interference mitigation in satellite
  communication systems.
\newblock Master's thesis, Louisiana State University, 2022.

\bibitem[CCSDS(2017)]{ccsds2017}
CCSDS.
\newblock Tm synchronization and channel coding—blue book.
\newblock \url{https://ccsds.org/Pubs/130x1g3e1.pdf}, September 2017.
\newblock CCSDS 131.0-B-3.

\bibitem[El-Khamy \& McEliece(2005)El-Khamy and McEliece]{elkhamy2005}
El-Khamy, M. and McEliece, R.~J.
\newblock Iterative algebraic soft-decision list decoding of reed--solomon
  codes.
\newblock arXiv preprint cs/0509097, September 2005.
\newblock URL \url{https://arxiv.org/abs/cs/0509097}.

\bibitem[{European Space Agency (ESA)}(2004)]{esa2004}
{European Space Agency (ESA)}.
\newblock Mars express -- the scientific payload.
\newblock ESA Special Publication SP 1240, August 2004.
\newblock URL \url{https://sci.esa.int/s/w0oQQlA}.

\bibitem[H.~Sawaguchi \& Wolf(2001)H.~Sawaguchi and Wolf]{Sawaguchi2001}
H.~Sawaguchi, S.~M. and Wolf, J.~K.
\newblock "a concatenated coding technique for partial response channels".
\newblock \emph{IEEE Transactions on Magnetics}, 37\penalty0 (2):\penalty0
  695--703, 2001.

\bibitem[Jiang \& Narayanan(2005)Jiang and Narayanan]{jiang2005}
Jiang, J. and Narayanan, K.~R.
\newblock Iterative soft input soft output decoding of reed--solomon codes by
  adapting the parity-check matrix.
\newblock arXiv preprint cs/0506073, June 2005.
\newblock URL \url{https://arxiv.org/abs/cs/0506073}.

\bibitem[Koetter \& Vardy(2003)Koetter and Vardy]{koetter2003}
Koetter, R. and Vardy, A.
\newblock Algebraic soft-decision decoding of reed--solomon codes.
\newblock \emph{IEEE Transactions on Information Theory}, 49\penalty0
  (11):\penalty0 2809--2825, 2003.
\newblock \url{https://www.site.uottawa.ca/~zhcheng/RS_soft_decision.pdf}.

\bibitem[McEliece \& Swanson(1986)McEliece and Swanson]{mceliece1986}
McEliece, R.~J. and Swanson, L.
\newblock On the decoder error probability for reed-solomon codes.
\newblock \emph{IEEE Transactions on Information Theory}, 32\penalty0
  (5):\penalty0 701--703, 1986.

\bibitem[Reed \& Solomon(1960)Reed and Solomon]{reed1960}
Reed, I.~S. and Solomon, G.
\newblock Polynomial codes over certain finite fields.
\newblock \emph{Journal of the Society for Industrial and Applied Mathematics},
  8\penalty0 (2):\penalty0 300--304, 1960.

\bibitem[Sun et~al.(2021)]{sun2021}
Sun, J. et~al.
\newblock Co-channel interference cancellation method based on deep neural
  network for leo satellite systems.
\newblock In \emph{China Satellite Navigation Conference (CSNC)}, volume 773 of
  \emph{Lecture Notes in Electrical Engineering}, pp.\  270--279. Springer,
  2021.
\newblock URL
  \url{https://link.springer.com/chapter/10.1007/978-981-16-3142-9_24}.

\bibitem[Townsend et~al.(2005)]{townsend2005}
Townsend, M.~J. et~al.
\newblock Leo download capacity analysis for a network of adaptive array ground
  stations.
\newblock Technical Report 20050210122, NASA, May 2005.

\bibitem[Vaswani et~al.(2017)Vaswani, Shazeer, Parmar, Uszkoreit, Jones, Gomez,
  Kaiser, and Polosukhin]{vaswani2017}
Vaswani, A., Shazeer, N., Parmar, N., Uszkoreit, J., Jones, L., Gomez, A.~N.,
  Kaiser, L., and Polosukhin, I.
\newblock Attention is all you need.
\newblock arXiv preprint arXiv:1706.03762, June 2017.
\newblock URL \url{https://arxiv.org/abs/1706.03762}.

\bibitem[Wesemeyer et~al.(2003)Wesemeyer, Sweeney, and Burgess]{wesemeyer2003}
Wesemeyer, S., Sweeney, P., and Burgess, D. R.~B.
\newblock Some soft-decision decoding algorithms for reed--solomon codes.
\newblock Technical report, Centre for Comm. Systems Research, University of
  Surrey, 2003.
\newblock \url{https://www.site.uottawa.ca/~zhcheng/RS_soft.pdf}.

\bibitem[Wicker \& Bhargava(1994)Wicker and Bhargava]{wicker1994}
Wicker, S.~B. and Bhargava, V.~K.
\newblock \emph{Reed-Solomon Codes and Their Applications}.
\newblock Wiley-IEEE Press, 1994.

\bibitem[Yi \& Zhong(2020)Yi and Zhong]{yi2020}
Yi, X. and Zhong, C.
\newblock Deep learning for joint channel estimation and signal detection in
  ofdm systems.
\newblock arXiv preprint arXiv:2008.03977, August 2020.
\newblock URL \url{https://arxiv.org/abs/2008.03977}.

\end{thebibliography}
\bibliographystyle{icml2025}

\end{document}